\documentclass[a4paper, 11pt]{article}
\usepackage{latexsym}
\title{On Noncommutative Classical Mechanics}
\vspace{2truecm}
\author{
A.E.F. DJEMA\"{I}\footnote{Permanent address : D\'epartement de
Physique, Facult\'e des Sciences, Universit\'e d'Oran Es--s\'enia,
31100, Oran, Algeria. Fax : (00).(213).(41).41.91.84,
e-mail : abedelfarid@yahoo.com} \\
\vspace{1truecm} \small{Abdus Salam International Centre for
Theoretical Physics,
34100, Trieste, Italy}\\
}
\vspace{1truecm}
\date{August, 2003}
\begin{document}
\maketitle
\begin{abstract}
In this work, I investigate the noncommutative Poisson algebra of
classical observables corresponding to a proposed general
Noncommutative Quantum Mechanics, \cite{1}. I treat some classical
systems with various potentials and some Physical interpretations
are given concerning the presence of noncommutativity at large
scales (Celeste Mechanics) directly tied to the one present at
small scales (Quantum Mechanics) and its possible relation with
UV/IR mixing.
\end{abstract}
\vspace{1truecm}
\begin{center}
\footnotesize{PACS : 02.40.Gh , 45.50Pk and 95.10.Ce }
\end{center}
\newpage
\section{Introduction:}
It is well-known that Quantum Mechanics (QM) can be viewed as a
noncommutative (matrix) symplectic geometry, \cite{2},
generalizing the usual description of Classical Mechanics (CM) as
a symplectic geometry. \\
In the context of the algebraic star--deformation theory, QM was
also described as a $\hbar$--deformation of the algebra
$\mathcal{A}_{0}$ of classical observables. The procedure consists
to replace the operator algebra issued from standard quantization
rules by the algebra $\mathcal{A}_{\hbar}$ of "quantum
observables" generated by the same classical observables obeying
actually a new internal law other than the usual point product,
the so--called Moyal star--product, \cite{3}, such that the
"classical" limit is guaranteed by $\hbar \rightarrow 0$. This is
the program
of "Quantization by deformation" carried out by Lichnerowicz and al.\\
Moreover, in the Lattice Quantum Phase Space, \cite{2,4}, the
discretization parameter $\frac{2\pi}{N}$ can be interpreted as a
deformation parameter. It is also well--known that, as the
"classical" limit $\hbar \rightarrow 0$ ensures the passage from
QM to CM, the passage, for instance, from Relativistic CM to
Non--relativistic CM is ensured by the "classical" limit $\beta =
\frac{v}{c}\rightarrow 0$, where $v$ is the velocity of the
classical particle and $c$ is the light velocity.\\
Recently, there has been a big interest in the study of various
Physical theories : String Theory,\cite{5}, Quantum Field
Theories, \cite{6}, QM,\cite{7}, Condensed Matter, \cite{8},
$\cdots$,
on noncommutative spaces.\\
Furthermore, the notion of \textbf{noncommutativity} may receive
different Physical interpretations. The most particular one
consists to do the parallel between the mechanics of a quantum
particle in the usual space in presence of a magnetic field $B$
and the mechanics of this quantum particle moving into a
noncommutative space. Furthermore, SUSY, through its
$Z_{2}$--graded algebra, may be viewed as a particular case of
noncommutativity. This means that \textit{superpartners} of
ordinary quantum particles can be studied only if one considers a
particular kind of noncommutativity, namely SUSY. Moreover, the
deformation parameter seems to be a \textit{fundamental constant}
which characterizes the Physics
described on a noncommutative space.\\
The aim of this work is, following the general formulation of
Noncommutative Quantum Mechanics (NCQM) proposed in \cite{1} and
generalizing the approach of \cite{9}, to discuss the associated
Noncommutative Classical Mechanics (NCCM)
and to treat some particular examples of classical potentials.\\
The work is organized as follows. Section $2$ is devoted to a
brief and methodic presentation of the passage from CM to QM, and
from QM to NCQM in view to fix notations. In section 3, I derive
the associated NCCM and discuss different aspects concerning the
star--deformed Poisson algebra and the resulting motion equations.
In section 4, I treat different cases of classical potentials
$V(x)$ like the free particle, the harmonic oscillator and, in
particular, the gravitational potential. The parallel between this
latter classical case and Coulomb potential in QM is discussed.
Finally, I devote section 5 to some conclusions and perspectives.
\section{CM $\rightarrow$ QM $\rightarrow$ NCQM}
Let us first start by considering a \textit{classical system} with
an Hamiltonian :
\begin{eqnarray}
H(x,p) = \frac{p^{2}}{2m} + V(x)   \label{3}
\end{eqnarray}
where the coordinates $x_{i}$ and the momenta $p_{i}$,
$i=1,\cdots,N$, generate the algebra $\mathcal{A}_{0}$ over the
Classical Phase Space (CPS) with the usual Poisson structure :
\begin{eqnarray*}
\left\{ x_{i} , x_{j} \right\}_{P} = 0~~~~~, ~~~~~\left\{ x_{i} ,
p_{j} \right\}_{P} = \delta_{ij}~~~~~,~~~~~\left\{ p_{i} , p_{j}
\right\}_{P} = 0
\end{eqnarray*}
or in terms of phase space variables $u_{a}$, $a=1,\cdots,2N$ :
\begin{eqnarray*}
\left\{ u_{a} , u_{b} \right\} = \omega_{ab}
\end{eqnarray*}
where $\omega$ is called the \textit{classical symplectic
structure} and is represented by the $2N \times 2N$ matrix :
\begin{eqnarray*}
\omega = \left(
\begin{array}{cc}
\textbf{0}               & \textbf{1}_{N \times N} \\
\textbf{-1}_{N \times N} & \textbf{0}
\end{array} \right)
\end{eqnarray*}
with $Det(\omega) = 1$.\\
Moreover, the motion equations of the classical system are given
by :
\begin{eqnarray*}
\dot{x}_{i} = \left\{ x_{i} , H \right\}~~~~~, ~~~~~ \dot{p}_{i} =
\left\{ p_{i} , H \right\} ~.
\end{eqnarray*}
Now, consider a \textbf{Dirac quantization} of this system :
\begin{eqnarray*}
\left\{ f , g \right\}_{P} \longrightarrow \frac{1}{i\hbar} \left[
\mathcal{O}_{f} , \mathcal{O}_{g} \right]
\end{eqnarray*}
where we denote by $\mathcal{O}_{f}$ the operator associated to a
classical observable $f$, with, in particular,
$\mathcal{O}_{x_{i}} = \textbf{x}_{i}$ and $\mathcal{O}_{p_{i}} =
\textbf{p}_{i}$. These operators generate the Heisenberg algebra :
\begin{eqnarray*}
\left[ \textbf{x}_{i} , \textbf{x}_{j} \right] =
\textbf{0}~~~~,~~~~\left[ \textbf{x}_{i} , \textbf{p}_{j} \right]
= i\hbar\delta_{ij} \textbf{1}~~~~,~~~~\left[ \textbf{p}_{i} ,
\textbf{p}_{j} \right] = \textbf{0} .
\end{eqnarray*}
Furthermore, the motion of this quantum system is governed by the
\textit{canonical equations} :
\begin{eqnarray*}
\dot{\textbf{x}}_{i} = \left[ \textbf{x}_{i} , \textbf{H}
\right]~~~~~, ~~~~~\dot{\textbf{p}}_{i} = \left[ \textbf{p}_{i} ,
\textbf{H} \right]
\end{eqnarray*}
where :
\begin{eqnarray*}
\textbf{H}(\textbf{x}, \textbf{p}) = \frac{\textbf{p}^{2}}{2m} +
V(\textbf{x})~.
\end{eqnarray*}
It is well known also that this \textbf{quantization} is
equivalent to a $\hbar$-star deformation of $\mathcal{A}_{0}$ such
that the Heisenberg operator algebra is replaced by the algebra
$\mathcal{A}_{\hbar}$ :
\begin{eqnarray}
\left\{ x_{i} , x_{j} \right\}_{\hbar} = 0~~~~,~~~~\left\{ x_{i} ,
p_{j} \right\}_{\hbar} = i\hbar \delta_{ij}~~~~,~~~~\left\{ p_{i}
, p_{j} \right\}_{\hbar} = 0  \label{1}
\end{eqnarray}
generated by the same classical observables but now obeying a
Moyal product :
\begin{eqnarray*}
\left( f \star_{\hbar} g \right) (u) = \exp \left[
\frac{i}{2}\hbar \omega^{ab} \partial^{(1)}_{a} \partial^{(2)}_{b}
\right] f(u_{1}) g(u_{2}) \vert_{u_{1} = u_{2} = u}
\end{eqnarray*}
where :
\begin{eqnarray*}
\omega^{ab} \omega_{bc} = \delta^{a}_{c}
\end{eqnarray*}
and
\begin{eqnarray*}
\left\{ f , g \right\}_{\hbar} = f \star_{\hbar} g - g
\star_{\hbar} f.
\end{eqnarray*}
Let us now consider another $\alpha$--star deformation of the
algebra $\mathcal{A}_{0}$, such that the internal law will be
characterized not only by the \textit{fundamental constant}
$\hbar$ but also by another deformation parameter (or more). This
can be performed by generalizing the usual symplectic structure
into another more general one, say $\alpha_{ab}$. For instance,
let us consider the algebra $\mathcal{A}_{\alpha}$ equipped with
the following star-product, \cite{1} :
\begin{eqnarray}
\left( f \star_{\hbar,\alpha} g \right)(u) = \exp\left[
\frac{i\hbar}{2} \alpha^{ab} \partial^{(1)}_{a} \partial^{(2)}_{b}
\right] f(u_{1}) g(u_{2}) \vert_{u_{1} = u_{2} = u}   \label{2}
\end{eqnarray}
such that
\begin{eqnarray*}
\alpha_{ab} = \left(
\begin{array}{cc}
\theta_{ij}                & \delta_{ij} + \sigma_{ij} \\
-\delta_{ij} - \sigma_{ij} & \beta_{ij}
\end{array} \right)
\end{eqnarray*}
where the $N \times N$ matrices $\theta$ and $\beta$ are assumed
to be antisymmetric :
\begin{eqnarray*}
\theta_{ij} = \epsilon_{ij}^{~~~k} \theta_{k}~~~~,~~~~\beta_{ij} =
\epsilon_{ij}^{~~~k}\beta_{k}
\end{eqnarray*}
while $\sigma$ is assumed to be symmetric and it will be
neglected since it is of second order, \cite{1}.\\
This new star-product generalizes the relations (\ref{1}) in the
following way :
\begin{eqnarray*}
\left\{ x_{i} , x_{j} \right\}_{\hbar,\alpha} = i\hbar
\theta_{ij}~~,~~\left\{ x_{i} , p_{j} \right\}_{\hbar,\alpha} =
i\hbar (\delta_{ij} + \sigma_{ij})~~,~~\left\{ p_{i} , p_{j}
\right\}_{\hbar,\alpha} = i\hbar \beta_{ij}
\end{eqnarray*}
and so gives rise to a NCQM defined by the following generalized
Heisenberg operator algebra :
\begin{eqnarray*}
\left[ \textbf{x}_{i} , \textbf{x}_{j} \right]_{\alpha} = i\hbar
\theta_{ij}\textbf{1}~~,~~\left[ \textbf{x}_{i} , \textbf{p}_{j}
\right]_{\alpha} = i\hbar (\delta{ij} + \sigma_{ij}
)\textbf{1}~~,~~\left[ \textbf{p}_{i} , \textbf{p}_{j}
\right]_{\alpha} = i\hbar \beta{ij}\textbf{1}. \label{5}
\end{eqnarray*}
In \cite{1}, we have found that the matrix $\sigma$ is tied to the
anticommutator of $\theta$ with $\beta$, and that the determinent
of the matrix $\alpha$ is given in function of $\rho = Tr(\theta
\beta) = Tr(\beta\theta)$. If we impose to the determinent to be
equal to $1$, then one obtains that :
\begin{eqnarray*}
\rho = - 2 \vec{\theta}.\vec{\beta}
\end{eqnarray*}
which is deeply linked to the Heisenberg Incertitude relations.
\section{Noncommutative Classical Mechanics}
The purpose of of this paper is precisely to study the
\textbf{noncommutative classical mechanics} which leads to the
NCQM as described in the previous section. The passage between
NCCM and NCQM is assumed to be realized via the following
\textbf{generalized Dirac quantization} :
\begin{eqnarray*}
\left\{ f , g \right\}_{\alpha}  \longrightarrow \frac{1}{i\hbar}
\left[ \mathcal{O}_{f} , \mathcal{O}_{g} \right]_{\alpha} .
\end{eqnarray*}
It follows that our \textbf{Noncommutative classical Mechanics} is
described by the $\alpha$-star deformed classical Poisson algebra
$\mathcal{A}_{\alpha}$ generated by the classical position and
momentum variables obeying to this new internal law, namely
(\ref{2}) without $i\hbar$ :
\begin{eqnarray*}
\left( f \star_{\alpha} g \right)(u) = \exp\left[ \frac{1}{2}
\alpha^{ab} \partial^{(1)}_{a} \partial^{(2)}_{b} \right] f(u_{1})
g(u_{2}) \vert_{u_{1} = u_{2} = u}
\end{eqnarray*}
such that :
\begin{eqnarray}
\left\{ x_{i} , x_{j} \right\}_{\alpha} = \theta_{ij}~~,~~\left\{
x_{i} , p_{j} \right\}_{\alpha} = \delta_{ij} +
\sigma_{ij}~~,~~\left\{ p_{i} , p_{j} \right\}_{\alpha} =
\beta_{ij}. \label{a}
\end{eqnarray}
Using the Hamiltonian (\ref{3}), we get the following Hamilton's
equations :
\begin{eqnarray}
\dot{x}_{i} = \left\{ x_{i} , H \right\}_{\alpha} &=&
\frac{p_{i}}{m} +  \theta_{ij} \frac{\partial V}{\partial x_{j}} +
\frac{1}{m} \sigma_{ij} p^{j} \simeq \frac{p_{i}}{m} + \theta_{ij}
\frac{\partial V}{\partial x_{j}} \label{b}\\
\dot{p}_{i}
=\left\{ p_{i} , H \right\}_{\alpha} &=& - \frac{\partial
V}{\partial x^{i}}  + \frac{1}{m} \beta_{ij} p^{j} - \sigma_{ij}
\frac{\partial V}{\partial x_{j}} \simeq - \frac{\partial
V}{\partial x^{i}}  + \frac{1}{m} \beta_{ij} p^{j}.\label{c}
\end{eqnarray}
In the noncommutative configuration space, the classical particle
obeys the following motion equations :
\begin{eqnarray}
m \ddot{x}_{i} &=& -\frac{\partial V}{\partial x^{i}} + m
\theta_{ij} \left( \frac{\partial^{2}V}{\partial x_{k} \partial
x_{j}} \right) \star \dot{x}_{k} \nonumber \\
               &+& \left[ (\textbf{1} + \sigma) \beta
(\textbf{1} + \sigma)^{-1} \right]_{ik} \dot{x}_{k}\nonumber \\
               &-& \left[
\sigma + (\textbf{1} + \sigma) \sigma + (\textbf{1} +
\sigma)\beta(\textbf{1} +\sigma)^{-1}\theta \right]_{ik} \left(
\frac{\partial V}{\partial x_{k}}\right) \nonumber \\
               &\simeq& -\frac{\partial V}{\partial x^{i}} +
\left[ m \theta_{ij} \left( \frac{\partial^{2}V}{\partial x_{k}
\partial x_{j}} \right)+ \beta_{ik} \right] \star \dot{x}_{k}\nonumber \\
               &+& O(\theta^{2}) + O(\beta^{2}) + O(\sigma). \label{4}
\end{eqnarray}
where $\textbf{1}$ means the $3 \times 3$ unit matrix.\\
The first term in the right side of this equation, that can be
obtained by taking the \textit{classical limit} ($\theta = \beta =
\textbf{0}$), represents the usual expression of a conservative
force which derives from a potential $V(x)$ present on the
commutative space (The second Newton law). The second term, which
has been found in \cite{9}, expresses a first correction to this
law depending on the presence of a noncommutativity only on the
configuration space ($\theta \neq \textbf{0}$ and $\beta =
\textbf{0}$) and also on the variations of the external potential
V(x), \cite{10}. The third term,  that is a kinetic correction
term, reflects a second correction due to the presence of a
noncommutativity only on the momentum sector of the classical
phase space ($\theta = \textbf{0}$
and $\beta\neq \textbf{0}$).\\
Hence, this result is very general in the sense that it takes into
account the noncommutativity on the whole phase space, since we
have shown in \cite{1}, that the presence of a noncommutativity on
the configuration space characterized by the parameter $\theta$
will automatically imply the presence of a noncommutativity on the
momentum sector characterized by the parameter $\beta$, such that
the two parameters are subject, through the parameter $\rho$, to a
lower bound constraint :
\begin{eqnarray*}
\rho = Tr[\theta\beta] = Tr[\beta\theta] = -2
\vec{\theta}.\vec{\beta} = -16.
\end{eqnarray*}
Indeed, the two parameters may exist and vary simultaneously but
are tied by the above constraint which has a direct Physical
interpretation (Heisenberg incertitude relations), \cite{1}.\\
Moreover, we remark that, in addition to the classical first term
in (\ref{4}), there is an additional term given in terms of
$\dot{x}_{k}$ that can be interpreted as the presence of some kind
of \textbf{viscosity} (resistivity) in the phase space due to its
noncommutativity property and also to the variations of the
potential.\\
Let us now consider a particular transformation on the usual
Classical Phase Space (CPS) that leads to the same results as of
the $\star_{\alpha}$--deformation on CPS, like the non--trivial
commutation relations (\ref{a}) or the motion equations (\ref{b}),
(\ref{c}) or (\ref{4}). Indeed, following the same approach as in
\cite{1}, we introduce the following transformation on usual CPS :
\begin{eqnarray}
x'_{i} = x_{i} - \frac{1}{2}\theta_{ij}p_{j}~~~~,~~~~p'_{i} =
p_{i} + \frac{1}{2}\beta_{ij}x_{j}   \label{e}
\end{eqnarray}
Firstly, it is easy to check that :
\begin{eqnarray}
\left\{ x'_{i} , x'_{j} \right\}_{P} = \theta_{ij}~~,~~\left\{
x'_{i} , p'_{j} \right\}_{P} = \delta_{ij} +
\sigma_{ij}~~,~~\left\{ p'_{i} , p'_{j} \right\}_{P} = \beta_{ij}.
\label{f}
\end{eqnarray}
where the symmetric $3\times3$--matrix $\sigma$ is given by :
\begin{eqnarray*}
\sigma = -\frac{1}{8} [ \theta \beta + \beta \theta ]   .
\end{eqnarray*}
Then, the usual Poisson brackets give the following Hamilton's
equations :
\begin{eqnarray}
\dot{x'}_{i} = \left\{ x'_{i} , H' \right\}_{P} &=&
\frac{p'_{i}}{m} + \theta_{ik} \frac{\partial V'}{\partial x'_{k}}
\\ \label{g}
\dot{p'}_{i} = \left\{ p'_{i} , H' \right\}_{P} &=& -
\frac{\partial V'}{\partial x'^{i}}  + \frac{1}{m} \beta_{ik}
p'^{k} .\label{h}
\end{eqnarray}
which looks like (\ref{b}) and (\ref{c}) respectively, and taking
care to
consider only first order terms in $\theta$ and/or $\beta$.\\
The motion equation on the usual configuration space is given by :
\begin{eqnarray}
m \ddot{x'}_{i} = -\frac{\partial V'}{\partial x'^{i}} + \left[ m
\theta_{ij} \left( \frac{\partial^{2}V'}{\partial x'_{k}
\partial x'_{j}} \right)+ \beta_{ik} \right] \dot{x'}_{k}
\label{i}
\end{eqnarray}
which looks also as the relation (\ref{4}), but now we are dealing
with commutative variables.
\section{Examples of classical systems}
Let us treat now some examples of classical systems : A free
particle ($V(x)=0$), an harmonic oscillator ($V(x) =
\frac{1}{2}Kx^{2}$) and a gravitational potential ($V(x) =
-\frac{K}{r}$).
\subsection{Free particle}
~~~~~In the case of a free classical particle described on the
noncommutative CPS, the motion equation (\ref{4}) reduces to :
\begin{eqnarray*}
m\ddot{x}_{i} = \beta_{ik}\dot{x}_{k} \Longrightarrow
m\vec{\gamma} = \vec{v} \land \vec{\beta} .
\end{eqnarray*}
This situation looks like the study of the motion of a classical
particle of charge $q$ described on the classical phase space in
presence of a magnetic field $\vec{B}$ :
\begin{eqnarray}
\vec{\beta} = q \vec{B} \label{6}
\end{eqnarray}
The quantum analog of this classical system behaves in the same
way, such that the gauge invariant velocity operator
$\vec{\textbf{v}}$ that defines the translation operator
$U(\vec{a}) = \exp{i\frac{m}{\hbar} \vec{a}.\vec{\textbf{v}}}$ on
the noncommutative configuration space do not commute in the sense
of (\ref{5}) :
\begin{eqnarray*}
\left[ \textbf{v}_{i} , \textbf{v}_{j} \right]_{\alpha} = i
\frac{\hbar}{m^{2}} \epsilon_{ij}^{~~k}\beta_{k}
\end{eqnarray*}
and do not associate :
\begin{eqnarray*}
\left[ \textbf{v}_{1} , \left[ \textbf{v}_{2} , \textbf{v}_{3}
\right]_{\alpha} \right]_{\alpha} + \left[ \textbf{v}_{3} , \left[
\textbf{v}_{1} , \textbf{v}_{2} \right]_{\alpha} \right]_{\alpha}
+ \left[ \textbf{v}_{2} , \left[ \textbf{v}_{3} , \textbf{v}_{1}
\right]_{\alpha} \right]_{\alpha} = \frac{\hbar^{2}}{m^{3}}
\vec{\nabla}. \vec{\beta} .
\end{eqnarray*}
This means that the quantum free particle of charge $q$ on a
noncommutative phase space looks like the well known quantum
mechanical problem of an ordinary quantum particle moving in the
configuration space in presence of a magnetic source,
specifically a magnetic monopole.\\
If we do the parallel between the two situations, then this will
lead to the interpretation of the presence of a noncommutative
perturbation on the phase space as a magnetic source (\ref{6}).\\
In this framework, the occurring of a nontrivial three cocycle
$\omega_{3}$, \cite{11} :
\begin{eqnarray*}
\omega_{3} = - \frac{1}{2\pi\hbar} \int d \vec{r}~
\vec{\nabla}.\vec{\beta}
\end{eqnarray*}
in the usual QM in presence of a magnetic source is deeply tied to
a certain topological perturbation of phase space since its
triangulation covering at very small scales means that the phase
space is no longer commutative, \cite{12}.\\
In the simple case where $\vec{\beta} = \beta \vec{k}$, which
means that the noncommutativity is present only on the plane
$(x,y)$, this implies a presence of magnetic field in the
direction of $z$--axis and so perturbs the $(x,y)$ plane.\\
However, within our framework, in the case of a free particle, we
have :
\begin{eqnarray*}
m\ddot{x'}_{i} = \beta_{ik}\dot{x'}_{k} = \epsilon_{i}^{~kl}
\dot{x}_{k}\beta_{l} = (\vec{v'} \wedge \vec{\beta})_{i} =
q(\vec{v'} \wedge \vec{B})_{i}
\end{eqnarray*}
We conclude that a free particle ($\ddot{x}_{i} = 0$) on the usual
CPS is now no longer free on the NCCPS. The noncommutativity on
CPS appears to be equivalent to the presence of some magnetic
field $\vec{B} =q^{-1}\vec{\beta}$.
\subsection{Harmonic oscillator}
Let us consider now the example of an harmonic oscillator
characterized by the potential :
\begin{eqnarray*}
V(x) = \frac{1}{2}k x^{2} = \frac{1}{2}k x_{i}\star_{\alpha} x^{i}
\end{eqnarray*}
In this case, the noncommutative Hamilton's equations (\ref{b})
and (\ref{c}) read :
\begin{eqnarray*}
\dot{x}_{i} = \frac{p_{i}}{m} +
k\theta_{ij}x_{j}~~~~,~~~~\dot{p}_{i} = -kx_{i} +\frac{1}{m}
\beta_{ij}p_{j}
\end{eqnarray*}
and the motion equations on the NC configuration space  become :
\begin{eqnarray*}
m\ddot{x}_{i} - \left[ \beta + mk \theta\right]_{ij} \dot{x}_{j} +
kx_{i} = 0
\end{eqnarray*}
or equivalently :
\begin{eqnarray*}
m\vec{\gamma} + \vec{\mu} \wedge \vec{v} + k \vec{x} = \vec{0}
\end{eqnarray*}
where
\begin{eqnarray*}
\vec{\mu} = mk\vec{\theta} + \vec{\beta}
\end{eqnarray*}
Investigating these motion equations, one finds that this
classical dynamical system on NC configuration space behaves like
a harmonic oscillator with the same frequency $\omega_{0} =
\sqrt{\frac{k}{m}}$, but in the plane perpendicular to the
direction of $\vec{\mu}$ :
\begin{eqnarray*}
\vec{\mu} . \left[ \vec{\gamma} + \omega^{2}_{0} \vec{x} \right] =
0
\end{eqnarray*}
Let's consider, for instance, the simple case where $\vec{\mu} =
\mu \vec{k}$, ($\theta_{1} =\theta_{2} = \beta_{1} = \beta_{2} =
0$ and $\mu = \mu_{3} = \beta_{3} + mk \theta_{3}$). Then, one has
:
\begin{eqnarray*}
\left\{ \begin{array}{l}
           m\ddot{x}_{1} + kx_{1} = \mu \dot{x}_{2} \\
           m\ddot{x}_{2} + kx_{2} = - \mu \dot{x}_{1}  \\
           m\ddot{x}_{3} + k x_{3} = 0
        \end{array} \right.
\end{eqnarray*}
The third equation confirms the fact that along the $z$--axis the
system still behaves as a harmonic oscillator with the same
frequency. Nevertheless, its motion in the $(x,y)$--plane is
governed by the two first mixed equations. Investigating these two
equations, we find :
\begin{eqnarray*}
\frac{1}{2}m [ \dot{x}_{1}\star_{\alpha}\dot{x}_{1} +
\dot{x}_{2}\star_{\alpha}\dot{x}_{2} ] + \frac{1}{2}k [
x_{1}\star_{\alpha}x_{1} + x_{2}\star_{\alpha}x_{2} ] =
\frac{1}{2}m v^{2} + \frac{1}{2}k r^{2} = H_{xy} = Constant.
\end{eqnarray*}
This looks like the expression of a conserved Hamiltonian of a
planar oscillator.\\
Finally, we conclude that, in this case, our 3D harmonic
oscillator on noncommutative CPS splits into two conservative
harmonic oscillators :
\begin{eqnarray*}
H = H_{xy} + H_{z}
\end{eqnarray*}
where
\begin{eqnarray*}
H_{z} = \frac{1}{2}m \dot{x}_{3}\star_{\alpha}\dot{x}_{3} +
\frac{1}{2}k x_{3} \star_{\alpha} x_{3}
\end{eqnarray*}
Let us now consider our approach based on considering the primed
commutative variables. In this case, the potential is given by :
\begin{eqnarray*}
V' = V(x') = \frac{1}{2}k x'^{2}
\end{eqnarray*}
and we can show that one obtains the same results as before.
Nevertheless, let us discuss the correction terms that occur in
the new Hamiltonian :
\begin{eqnarray*}
H' = H - \frac{1}{2m} \vec{L} . \vec{\mu}
\end{eqnarray*}
This confirms the fact that our 3D harmonic oscillator on
noncommutative CPS is equivalent to the usual 3D harmonic
oscillator of charge $q$ in presence of some magnetic field :
\begin{eqnarray*}
\vec{B} = q^{-1} \vec{\mu}
\end{eqnarray*}
\subsection{Gravitational potential}
Let's consider a particle of mass $m$ and charge $q$ moving in a
gravitational potential :
\begin{eqnarray*}
V(r) = - \frac{k}{r}
\end{eqnarray*}
where $r = \sqrt{x_{i} \star_{\alpha} x^{i}}$. Let's set :
\begin{eqnarray*}
\Omega_{i} = \frac{k}{r^{3}}\theta_{i}
\end{eqnarray*}
which we will call the \textit{angular velocity}. Then, the NC
Hamilton's equations read :
\begin{eqnarray*}
\dot{x}_{i} &=& \frac{p_{i}}{m} +
\theta_{ij}\frac{kx^{j}}{r^{3}} = \frac{p_{i}}{m} +
(\vec{x} \wedge \vec{\Omega} )_{i}\\
\dot{p}_{i} &=& - \frac{kx_{i}}{r^{3}} +
\frac{1}{m}\beta_{ij}p^{j} = - \frac{kx_{i}}{r^{3}} + \frac{1}{m}
(\vec{p} \wedge \vec{\beta})_{i}
\end{eqnarray*}
and the motion equations on the NC configuration space become :
\begin{eqnarray*}
m\ddot{x}_{i} = - \frac{x_{i}}{r} \frac{k}{r^{2}} +
m\epsilon_{i}^{~jk}\left( \dot{x}_{j} \Omega_{k} + x_{j}
\dot{\Omega}_{k} \right) + \epsilon_{i}^{~jk}\dot{x}_{j} \beta_{k}
\end{eqnarray*}
or equivalently :
\begin{eqnarray}
m\vec{\gamma} = - \frac{k}{r^{2}} \frac{\vec{x}}{r} + m \left(
\vec{\dot{x}} \wedge \vec{\Omega} + \vec{x} \wedge
\vec{\dot{\Omega}}\right) + \vec{\dot{x}} \wedge \vec{\beta} = -
\frac{k}{r^{2}} \frac{\vec{x}}{r} + \vec{\dot{x}} \wedge
\vec{\sigma} + \vec{x} \wedge \vec{\dot{\sigma}} \label{8}
\end{eqnarray}
where
\begin{eqnarray*}
\vec{\sigma} = \vec{\beta} + \frac{km}{r^{3}}\vec{\theta} =
\vec{\beta} + m \vec{\Omega}
\end{eqnarray*}
These equations of motion are different from the ones obtained in
\cite{9} by a term that comes from the noncommutativity parameter
$\beta$ which is not considered there.\\
Moreover, it is straightforward to check that the Hamiltonian is a
constant of motion :
\begin{eqnarray*}
\dot{H} = \frac{1}{2m} \left[ \dot{p}_{i} \star_{\alpha} p^{i} +
p_{i} \star_{\alpha} \dot{p}^{i} \right] + \dot{V}(r) =
\frac{1}{m}p_{i}\dot{p}^{i} + \frac{k}{r^{3}}x_{i}\dot{x}^{i} = 0
\end{eqnarray*}
and that the components of the angular momentum of this system on
NCCPS are no longer conserved :
\begin{eqnarray*}
L^{NC}_{i} = \epsilon_{i}^{~jk}x_{j} \star p_{k} = L^{C}_{i} -
\frac{mk}{r^{3}}\left[ \vec{x} \wedge (\vec{x} \wedge
\vec{\theta}) \right]_{i} = L^{C}_{i} - m \left[ \vec{x} \wedge
(\vec{x} \wedge \vec{\Omega}) \right]_{i}
\end{eqnarray*}
where :
\begin{eqnarray*}
L^{C}_{i} = \epsilon_{i}^{~jk}x_{j}(m\dot{x}_{k})
\end{eqnarray*}
is the conserved angular momentum on usual CPS.\\
Nevertheless, the component along the $\vec{\sigma}$ axis of the
angular momentum is conserved :
\begin{eqnarray*}
\vec{L}^{NC}.\vec{\sigma} = \vec{L}^{C}.\vec{\sigma} =
\epsilon^{ijk} \sigma_{i} x_{j}(m\dot{x}_{k})
\end{eqnarray*}
In another hand, we remark from (\ref{8}), that relatively to the
$\vec{\sigma}$ axis our system still remains "classical", i.e. :
\begin{eqnarray}
m\vec{\gamma}.\vec{\sigma} =
-\frac{k}{r^{2}}\frac{\vec{x}.\vec{\sigma}}{r} \label{9}
\end{eqnarray}
Then, it is more indicated to study the motion of the system in
the plane perpendicular to the $\vec{\sigma}$ axis. For this
reason, in the following we will consider only one independent
noncommutative parameter, namely $\sigma = \sigma_{3} = \beta + m
\Omega$, with $\theta_{1} = \theta_{2} = \beta_{1} = \beta_{2} =
0$ and $\theta = \theta_{3}$, $\beta = \beta_{3}$, $\Omega =
\Omega_{3}$. Firstly, along the $\vec{\sigma}$ axis the motion of
our system is governed by (See (\ref{9})) :
\begin{eqnarray}
m\ddot{x}_{3} = - \frac{\partial V}{\partial x_{3}} = -
\frac{kx_{3}}{r^{3}}  \label{15}
\end{eqnarray}
Now, let us express the motion equations (\ref{8})of this system
on the $(x,y)$-plane in terms of polar coordinates $(\rho, \phi)$
:
\begin{eqnarray}
\left\{ \begin{array}{ll} m [ \ddot{\rho} - \rho \dot{\phi}^{2} ]
= - \frac{\partial V(\rho)}{\partial \rho} + m\rho\sigma\dot{\phi}
= - \frac{k}{\rho^{2}} + m\rho\dot{\phi}\Omega +
\rho\dot{\phi}\beta \label{10}  \\
\frac{d}{dt}\left[ m \rho^{2}
\dot{\phi} \right] = -\rho \frac{d}{dt} \left( \rho \sigma \right)
= -m\rho\frac{d}{dt}\left( \rho \Omega \right) - \beta \rho
\dot{\rho} \end{array} \right. \label{11}
\end{eqnarray}
where we have considered the case of equatorial orbits $(\varphi =
\frac{\pi}{2} \Longrightarrow r = \rho)$.\\
It is easy to check from (\ref{11}), that the quantity :
\begin{eqnarray*}
M = \rho^{2} ( m\dot{\phi} + \sigma ) - m\theta V -
\frac{\beta}{2}\rho^{2} = m\rho^{2} \dot{\phi} +
\frac{2mk\theta}{\rho} + \frac{\beta}{2}\rho^{2}
\end{eqnarray*}
is a constant of motion since $\dot{M} = 0$.\\
Returning to the equation (\ref{10}), we find :
\begin{eqnarray*}
m\ddot{\rho} + \frac{k}{\rho^{2}} - \frac{M^{2}}{m\rho^{3}} +
\frac{3kM\theta}{\rho^{4}} = 0
\end{eqnarray*}
where we have neglected second order terms in $\theta$ and
$\beta$.\\
In order to deduce the trajectory equation $\rho =
\rho(\phi)$, let us introduce the following change :
\begin{eqnarray*}
u = \frac{1}{\rho}
\end{eqnarray*}
Then, we obtain the following differential equation  :
\begin{eqnarray}
\left[ Mu^{3} - 4km\theta u^{4} - \beta u \right] \left(
\frac{d^{2}u}{d\phi^{2}}\right) - \left[ 2km\theta u^{3} + \beta
\right]\left( \frac{du}{d\phi}\right)^{2} - k\frac{m}{M}u^{3} +
Mu^{4} - 3km\theta u^{5} = 0 \label{12}
\end{eqnarray}
that differs from the one obtained in \cite{9} by additional terms
in $\beta$ and missing terms of second order in $\theta$ and
$\beta$.\\
In the classical case, i.e. at the zero order ($\theta = \beta =
0$), we obtain the ordinary Kepler motion equation :
\begin{eqnarray*}
\frac{d^{2}u_{0}}{d\phi^{2}} + u_{0} = \frac{1}{b}
\end{eqnarray*}
where
\begin{eqnarray*}
b = \frac{M^{2}}{km}.
\end{eqnarray*}
The solution of this equation is given by the elliptic trajectory
:
\begin{eqnarray*}
u_{0} = \frac{1+e \cos\phi}{b}
\end{eqnarray*}
where $e$ is some parameter representing the eccentricity of the ellipse. \\
At first order in $\theta$ and $\beta$, we propose the following
solution :
\begin{eqnarray}
u = u_{0} + \theta u_{1} + \beta u_{2} \label{16}
\end{eqnarray}
Replacing in (\ref{12}), one obtains the following differential
equations :
\begin{eqnarray*}
\left\{ \begin{array}{ll}
        \frac{d^{2}u_{1}}{d\phi^{2}} + u_{1} = F_{1}(\phi)  \label{13}\\
        \frac{d^{2}u_{2}}{d\phi^{2}} + u_{2} = F_{2}(\phi)  \label{14}
         \end{array} \right.
\end{eqnarray*}
where
\begin{eqnarray*}
F_{1} &=& \frac{M}{b^{3}}\left[ 2e\cos(\phi) -
\frac{3}{2}e^{2}\cos(2\phi) + \frac{e^{2}+6}{2} \right] \\
F_{2} &=& - \frac{be}{M}\left[ \frac{\cos(\phi) + e
\cos(2\phi)}{\left( 1 + e \cos(\phi)\right)^{3}} \right]
\end{eqnarray*}
The first differential equation admits the following general
solution :
\begin{eqnarray*}
u_{1}(\phi) = \frac{M}{b^{3}} \left[ e\phi \sin(\phi) +
\frac{e^{2}}{2}\cos(2\phi) + \frac{e^{2}+6}{2}\right]
\end{eqnarray*}
while the second one admits a more complicated general solution
which looks like :
\begin{eqnarray*}
u_{2} &=& - \frac{be}{M}\sin(\phi)\left\{ A_{0}\phi\sin(\phi) +
A_{1} \mbox{arctanh} \left[ a \tan(\frac{\phi}{2}) \right] +
A_{2}\cot(\phi) + A_{3}\csc(\phi) +
A_{4}\cot\left(\frac{\phi}{2}\right) \right. \\
     &+& \left.
A_{5} \tan\left(\frac{\phi}{2}\right) + A_{6}\cot(\phi)\ln[bu_{0}]
+ A_{7}\left(\frac{A_{8}\sin(\phi) +
A_{9}\sin(2\phi)}{b^{2}u^{2}_{0}}\right) \right\}
\end{eqnarray*}
where $A_{0} = -\frac{2}{e^{2}}$ and the other coefficients are
functions of $e$.\\
Then, to first order in $\theta$ and $\beta$, the general solution
of (\ref{12}) is given by (\ref{16}), i.e. :
\begin{eqnarray*}
u &=& u_{0} + \theta u_{1} + \beta u_{2} = \frac{1 +
e\cos(\phi)}{b} + \left[ \frac{Me}{b^{3}}\theta +
\frac{2b}{Me}\beta \right] \phi\sin(\phi) + \theta \left[
....\right] + \beta \left[ ....\right] \\
&\approx& \left[ \frac{1 + e\cos\left[\phi (1 - \frac{\xi}{b})
\right]}{b}\right] + .....
\end{eqnarray*}
The remarkable point is the appearance of terms linear in $\phi$
in the perturbation terms $u_{1}$ and $u_{2}$. These interesting
terms, that let the original ellipse $u_{0}$ change when it
precesses, permit us to calculate the possible perihelion shift
per revolution due to noncommutativity :
\begin{eqnarray*}
\delta \phi_{NC} = 2\pi \left[ \frac{\xi}{b}\right]
\end{eqnarray*}
where :
\begin{eqnarray*}
\xi = \frac{M}{b}\theta + \frac{2b^{3}}{Me^{2}}\beta
\end{eqnarray*}
Taking into account that :
\begin{eqnarray*}
k = mm_{s}G~~~~~~,~~~~~~b = a(1-e^{2})
\end{eqnarray*}
where $m_{s}$ is the sun mass and "$a$" is the average radius of
the ellipse, then :
\begin{eqnarray*}
\delta \phi_{NC} = 2\pi \left\{\frac{M}{b^{2}}\theta +
\frac{2b^{2}}{Me^{2}}\beta \right\} = 2\pi \left\{
\kappa^{\frac{1}{2}}\theta +
\frac{2}{e^{2}}\kappa^{-\frac{1}{2}}\beta \right\}
\end{eqnarray*}
with
\begin{eqnarray*}
\kappa = \frac{m^{2}m_{s}G}{a^{3}(1-e^{2})^{3}}
\end{eqnarray*}
Furthermore, it has been shown that in the context of General
Relativity, the advance of the perihelion with the Schwarzschild
metric is given by , \cite{13} :
\begin{eqnarray*}
\delta \phi_{RG} = 2\pi \left\{
\frac{3m_{s}G}{c^{2}a(1-e^{2})}\right\}
\end{eqnarray*}
Then, it follows that :
\begin{eqnarray*}
\delta \phi_{NC} = \lambda \delta \phi_{RG}
\end{eqnarray*}
where
\begin{eqnarray*}
\lambda = \frac{a(1-e^{2})c^{2}}{3Gm_{s}}\left[
\kappa^{\frac{1}{2}} \theta + \frac{2}{e^{2}}\kappa^{-\frac{1}{2}}
\beta \right]
\end{eqnarray*}
In the particular case of the Mercury planet, and using the
following data :
\begin{eqnarray*}
a &\approx& 6 \times 10^{10}m~~,~~e \approx 0,2~~,~~m \approx 3,3
\times 10^{23}kg \\
m_{s} &\approx& 2\times 10^{30}kg~~,~~G \approx 7 \times
10^{-11}m^{3}kg^{-1}s^{-2}~~,~~\hbar \approx 6,6 \times 10^{-34}
Js
\end{eqnarray*}
we found that:
\begin{eqnarray*}
\kappa \approx 10^{34}~~kg^{2}/s^{2}~~~~,~~~~\lambda \approx 1.2
\times 10^{7} \left[ 10^{17}\theta  + 50 \times 10^{-17} \beta
\right]
\end{eqnarray*}
and then, the perihelion shift is of order :
\begin{eqnarray*}
\delta \phi_{NC} \approx 2\pi \left[ 10^{17} \theta + 50 \times
10^{-17}\beta \right]
\end{eqnarray*}
Let us recall that the parameters $\theta$ and $\beta$ have been
at first considered as perturbation parameters, so they are very
small, \cite{1}. Then, from the above relation, one can deduce
that the contribution of the second parameter is very small
compared to the one of the first parameter. So, we can ignore it.
In this case, our results will be very close to those obtained in
\cite{9}. In fact, let us evaluate an order of the first parameter
by comparing $\delta \phi_{NC}$ to the experimental data. \\
Knowing that the observed perihelion shift for Mercury is,
\cite{13} :
\begin{eqnarray*}
\delta\phi_{obs} = 2\pi (7.98734 \pm 0.0003)\times 10^{-8} rad/rev
\end{eqnarray*}
and assuming that $\delta \phi_{NC} \approx \delta \phi_{obs}$, it
follows that :
\begin{eqnarray*}
\theta \approx 8 \times 10^{-25} s/kg
\end{eqnarray*}
Now, since the noncommutativity effect is considered as a quantum
effect of gravity, \cite{12}, let us calculate :
\begin{eqnarray*}
\sqrt{\hbar \theta} \approx 23 \times 10^{-30} m.
\end{eqnarray*}
Moreover, General Relativity predicts for the perihelion shift :
\begin{eqnarray*}
\delta\phi_{RG} = 2\pi (7.987344)\times 10^{-8} rad/rev
\end{eqnarray*}
So, we can evaluate a lower bound for $\theta$ by means of the
difference between the General relativity prediction of the shift
and the observed one :
\begin{eqnarray*}
\mid \delta\phi_{NC}\mid \leq \mid \delta\phi_{GR} -
\delta\phi_{obs}\mid \approx 4 \times 10^{-14}
\end{eqnarray*}
Then, we get :
\begin{eqnarray*}
\theta \leq 6 \times 10^{-32} &\rightarrow& \hbar \theta \leq 40
\times 10^{-62} m^{2}\\
&\rightarrow& \sqrt{\hbar \theta} \leq 63 \times 10^{-32} m
\approx (4 \times 10^{4})L_{P} \\
&\rightarrow& \frac{1}{\sqrt{\hbar \theta}}\geq 1.6 \times 10^{30}
m^{-1}
\end{eqnarray*}
where $L_{P}$ represents the Planck scale.\\
Now, let us return to our approach that makes use of primed
variables (\ref{e}). In this framework, the Hamiltonian of our
system on NCCPS reads as :
\begin{eqnarray}
H' = H - \frac{1}{2m} {\vec{L}}^{C}.\vec{\sigma}  \label{7}
\end{eqnarray}
From (\ref{7}), we can interpret the manifestation of
noncommutativity on CPS as being equivalent to the presence of
some "magnetic field" $\vec{B} = q^{-1}\vec{\sigma}$ that
interacts with our system of charge $q$.\\
In this framework, the components of the angular momentum on NCCPS
are given by :
\begin{eqnarray*}
L^{NC}_{i} = L'_{i} = \epsilon_{i}^{~jk}x'_{j}p'_{k} = L^{C}_{i} +
\frac{1}{2} \left[ \vec{x} \wedge (\vec{x} \wedge \vec{\beta}) -
(\vec{p} \wedge \vec{\theta}) \wedge \vec{p} \right]_{i}
\end{eqnarray*}
Moreover, it is easy to see that following our framework, we will
obtain nearly the same results as described before.\\
\section{Conclusion}
In this work, I have studied the noncommutative classical
mechanics related to the Noncommutative Quantum Mechanics as
described in \cite{1}. The same interpretations have been given to
the occurrence of noncommutativity effects as in the quantum case.
Treating the particular case of a gravitational potential, which
is relevant at large scales and which looks like the Coulomb
potential at small scales, I show that there is a correction to
the perihelion shift of Mercury, and with a parameter $\hbar
\theta$ of the order of $10^{-56} m^{2}$ we are in presence of an
observable deviation. \\
Let us remark that the second NC parameter $\beta$ does not
contribute to this correction compared to the contribution of the
parameter $\theta$.\\
Finally, the main point in our work is the fact that the NC
parameters which are initially present at a quantum level, occur
also at large scales. So, there is a deep link between Physics at
small scales and Physics at large scales as it is predicted by
UV/IR mixing. This confirm the results obtained in \cite{9}.\\

\textbf{Acknowledgments}: I would like to thank Arab Fund and the
Associateship scheme of Abdus Salam ICTP for their support and
help. I would also like to thank E. Yuzbashyan for useful
discussions.


\begin{thebibliography}{99}
\bibitem{1} ~A.E.F. Djemai \& H. Smail : \textit{On Quantum Mechanics
             on Noncommutative Quantum Phase Space}, hep-th/0309006, Submitted
             for publication to Comm. Theor. Phys.
\bibitem{2}~A.E.F. Djemai : Int. J. Theor. Phys. \textbf{35}(3),
            (1996)519
\bibitem{3}~J.E. Moyal : Proceedings of the Cambridge Philosophical
Society \textbf{45} (1949)99 ;
            J. Vey : commentari Mathematici Helvetici \textbf{50} (1975)421 \\
            M. Flato \& al : Compositio Mathematica \textbf{31}
            (1975)47 ;
            M. Flato \& al : J. Math. Phys. \textbf{17} (1976)1754
            ;
            F. Bayen \& al : Lett. Math. Phys. \textbf{1}
            (1977)521 ;
            F. Bayen \& al : Annals of Physics \textbf{111}
            (1978)61, 111
\bibitem{4}~A.E.F. Djemai : Int. J. Mod. Phys. \textbf{A10},
             (1995)3303
\bibitem{5}~~A. Connes \& al : JHEP \textbf{9802} (1998)003
[hep-th/9711162] ;
              M.R. Douglas and C.M. Hull :~JHEP \textbf{9802}
              (1998)008  [hep-th/9711165] ;
              C. Chu and P. Ho : Nucl. Phys.
              B\textbf{550}(1999)151 [hep-th/9812219]; Nucl. Phys.
              B\textbf{568} (2000)447 [hep-th/9906192] ;
              V. Schomerus : JHEP \textbf{9906} (1999)030
              [hep-th/9903205] ;
              N. Seiberg and E. Witten : JHEP \textbf{9909} (1999)032 [hep-th/9908142] and references therein.
\bibitem{6}~M. Chaichian \& al : Phys. Rev. Lett.
             \textbf{86}(2001)2716 [hep-th/0010175]; \textit{Comments on the
             Hydrogen Atom in the Noncommutative Space}
             [hep-th/0212259] ; R. Szabo : \textit{Quantum Field
             Theory on Noncommutative Spaces} [hep-th/0109162]
\bibitem{7}~P-M. Ho and H-C. Kao : \textit{Noncommutative Quantum Mechanics from Noncommutative Quantum Field
            Theory} [hep-th/0110191] ; D. Kochan and M. Demetrian : \textit{QM
            on Noncommutative Plane} [hep-th/0102050] ; J. Gamboa \& al : Phys. Rev. D\textbf{64} (2001)267
              [hep-th0010220] ;
              V.P. Nair and A.P. Polychronakos : Phys. Lett. B
              \textbf{505} (2001)267 [hep-th/0011172] ;
              B. Morariu and A.P. Polychronakos : Nucl. Phys.
              B\textbf{610} (2001)531 [hep-th/0102157] ;
              J. Gamboa \& al : \textit{Noncommutative Quantum Mechanics : The Two--dimensional Central
              Field} [hep-th/0106125] ;
              S. Bellucci \& al : \textit{Two phases of the Noncommutative Quantum
              Mechanics} [hep-th/0106138] ;
              H.R. Christiansen and F.A. Schaposnik : \textit{Noncommutative Quantum Mechanics and Rotating
              Frames} [hep-th/0106181] ;
              C. Acatrinei : \textit{Path Integral Formulation of Noncommutative Quantum
              Mechanics} JHEP \textbf{0109} (2001)007 [hep-th/0107078]
\bibitem{8}~Z.F. Ezawa : \textit{Quantum Hall effects : Field
Theoretical Approach and Related Topics}, World Scientific, (2000)
\bibitem{9}~J.M. Romero \& J.D. Vergara : \textit{The Kepler Problem and Non
Commutativity} [hep-th/0303064] ; B. Mirza \& M. Dehghani :
\textit{Noncommutative Geometry and the Classical Orbits of
Particles in a Central Force potential} [hep-th/0211190] ;
\bibitem{10}~J.M. Romero \& al : \textit{Newton's Second Law in a
Noncommutative Space}, [hep-th/0211165]
\bibitem{11}~R. Jackiw : \textit{Topological Investigations of Quantized Gauge
Theories} in \textit{Current Algebra and Anomalies} , World
Scientific (1985)317
\bibitem{12}~H.S. Snyder : Phys. Rev. \textbf{71} (1946)38 ;
              C.N. Yang : Phys. Rev. \textbf{72} (1947)874.
\bibitem{13}~S. Pireaux \& al : \textit{Solar Quadrupole Moment and Purely Relativistic Gravitation Contributions
             to Mercury's perihelion Advance}, [astro-ph/0109032]
\end{thebibliography}
\end{document}